\documentclass[12pt,preprint]{aastex}
\usepackage{emulateapj5}                                                                                     

\newcommand{\lya}{Ly$\alpha$}

\newcommand{\etal}{et al.}

\slugcomment{Submitted to the Astronomical Journal}

\shorttitle{VLT observations of the z=6.28 quasar SDSS 1030+0524}
\shortauthors{L. Pentericci, et al.}

\begin{document}


\title{VLT optical and near-IR observations of the z=6.28 quasar SDSS 1030+0524 }


\author{Laura Pentericci\altaffilmark{\ref{MPIA}}, 
Xiaohui Fan\altaffilmark{\ref{IAS}}, 
Hans Walter Rix\altaffilmark{\ref{MPIA}}, 
Michael A. Strauss\altaffilmark{\ref{Princeton}},
Vijay K. Narayanan\altaffilmark{\ref{Princeton}},
Gordon T. Richards\altaffilmark{\ref{PSU}},
Donald P. Schneider\altaffilmark{\ref{PSU}},      
Julian Krolik\altaffilmark{\ref{JH}}, 
Tim Heckman\altaffilmark{\ref{JH}},
Jonathan Brinkmann\altaffilmark{\ref{APO}},
Don Q. Lamb\altaffilmark{\ref{Chicago}},
Gyula P. Szokoly\altaffilmark{\ref{AIP}}
}

\altaffiltext{1}{Based on observations carried out at the European
Southern Observatory, Paranal, Chile as part of the ESO DDT observing programs 
  267.A-5689 and 267.A-5696}
\altaffiltext{2}{Max Planck Institut f\"ur Astronomie, K\"{o}nigstuhl 17, D-69171 Heidelberg, Germany\label{MPIA}}
\altaffiltext{3}{Institute for Advanced Study, Olden Lane,
Princeton, NJ 08540 USA \label{IAS}}    
\altaffiltext{4}{Princeton University Observatory, Princeton,
 08544 USA \label{Princeton}}  
\altaffiltext{5}{Department of Astronomy and Astrophysics,
The Pennsylvania State University,
University Park, PA 16802 USA
\label{PSU}}
\altaffiltext{6}{Department of Physics and Astronomy, Johns Hopkins University, Baltimore, MD 21218  USA \label{JH}}
\altaffiltext{7}{Apache Point Observatory, P. O. Box 59,
Sunspot, NM 88349-0059 USA     \label{APO}}
\altaffiltext{8}{University of Chicago, Astronomy \& Astrophysics
Center, 5640 S. Ellis Ave., Chicago, IL 60637 USA  \label{Chicago}}
\altaffiltext{9}{Astrophysikalisches Institut Potsdam, An der Sternwarte 16, 14482 Postdam, Germany \label{AIP}}



\begin{abstract}
We present new VLT spectroscopic observations of the most distant quasar known, SDSS J1030+0524
at z$=$6.28, which was recently discovered by the Sloan Digital Sky Survey.
We confirm the presence of a complete Gunn-Peterson trough caused by neutral hydrogen in the intergalactic medium. 
There is no detectable flux over the wavelength range from 8450 to 8710 \AA.
We set a stronger limit on the drop of the flux level blueward of the Ly$\alpha$ line: a factor of $>$ 200.
Below 8450 \AA \ the spectrum shows a rise in flux, with a large fraction 
($>$ 60 $\%$) of the total emission 
produced by few narrow features of transmitted flux.    
We discuss the proximity effect around this quasar, 
with the presence of transmitted flux with many absorption features in a region of about 23$h^{-1}$ comoving Mpc.
If assuming the surrounding medium were completely neutral, 
the size of this region would imply a 
 quasar lifetime of $\sim$ 1.3 $\times$10$^7$ years.

We also present near-IR spectroscopy of both SDSS J1030+0524 and of  
SDSS J1306+05, the second most distant quasar known, at redshift 6.0. 
We combine measurements of the CIV line and 
limits on the HeII emission from the near-IR spectra
with the NV line measurements from the optical spectra to derive the metal abundances of 
these early quasar environments.
The results are indistinguishable from those of lower redshift quasars
and indicate little or no evolution in the metal abundances 
from z$\sim$ 6 to z$\sim$2.
The line ratios suggest supersolar metallicities, 
implying that the first stars around the quasars must have formed at least a few hundreds of Myrs prior to the observation, i.e., at redshifts higher than 8. 
\end{abstract}

\keywords{
cosmology: observations ---
galaxies: formation ---
galaxies: quasars: absorption lines ---
galaxies: quasars: individual (SDSS 1030+0524, SDSS 1306+0356)
}  


\section{Introduction}
Quasars are amongst the most luminous objects in the Universe,
allowing  us to study them and any intervening material out to
 very large distances, corresponding to
look-back times when the Universe was very young.  Hence finding and
studying quasars at high redshifts is one of the
best ways to constrain the physical conditions in the early Universe.
The mere existence of luminous quasars at such early times, and the implied presence of black holes with M$\ \ge 10^9$ M$_\odot$ 
place stringent limits on the epoch
at which massive condensed structures formed, thereby  constraining structure
formation models (e.g.\ Efstathiou and Rees 1988).  
 At high redshifts such luminous quasars must be associated with very massive halos, hence they are expected to be found near high peaks (4-5 $\sigma$ or more) of the large scale
density field
(e.g.\ Efstathiou \& Rees 1988, Nusser \& Silk 1993) that collapsed sufficiently
early.
Furthermore, several arguments suggest that the formation of elliptical
galaxies and massive bulges  are paralleled by an early quasar phase
(e.g.\ Kauffmann \& Haehnelt 2000).

The spectra of high redshift quasars  contain important
information on the enrichment history of the gas in the quasar
environment, and probe the star formation preceding the epoch at which
the quasars are observed,
possibly the first stars that formed in massive collapsed structures
(e.g.\ Hamann \& Ferland 1999). 
Finally,  high redshift quasars serve as probes 
of the intergalactic medium via the absorption of the Lyman $\alpha$ forest.

The Sloan Digital Sky Survey (SDSS -- York et al. 2000) has, amongst its scientific aims, the construction of the largest 
sample of quasars ever, with more than 10$^5$ objects 
spanning a large range of redshift and luminosities.
The SDSS has already found an unprecedented number of new high redshift 
quasars,
including more than 200 new quasars at z$\ge$4 (e.g. Fan \etal\ 2000, Zheng \etal\ 2000, Anderson \etal\ 2001, Fan et al. 2001a, Schneider \etal\ 2001). 
These high redshift quasars have been efficiently 
selected by their distinctive position in
 color-color diagrams, with characteristic colors due
to the main feature of the quasar
 spectra, viz., the strong Ly$\alpha$ emission line,
the Ly$\alpha$ forest and the Lyman limit.  

Recently, Fan et al.\ (2001b) presented the discovery of 
SDSS J1030+0524, found  during 
 follow-up spectroscopy of $i$-band drop-out
objects, i.e. objects showing very red $i^* - z^*$ color and relatively blue 
$z^*$-J colors (following the notation of previous SDSS papers, we use the superscript $^*$ 
for the photometry, and the letters alone for the filters, Stoughton \etal\ 2002).
Moderate signal to noise ratio (S/N) optical spectra taken with the
ARC 3.5m telescope
showed very strong (rest-frame equivalent width EW$\sim$50 \AA)
Ly$\alpha$ + NV emission at $\sim$8850 \AA \ and $\sim$9050 \AA, \ 
respectively.
Based on a fit of the CIV line the redshift was estimated estimated 
to be  z$\sim$ 6.28 which makes it unambiguously the most distant quasar known.
Follow up optical spectroscopic observations with Keck (Becker et al. 2001)
have shown the first  clear detection of a complete Gunn-Peterson trough 
(Shklovsky 1964, Scheuer 1965, Gunn \& Peterson 1965)
in the spectrum of this quasar. The flux level drops by a factor of at 
least 140 relative to an estimate of the unabsorbed continuum level,
and the spectrum is consistent with zero 
flux in the Ly$\alpha$ forest region immediately 
blueward of the Ly$\alpha$ emission line. 
Even if the existence of the trough itself does
not imply that the quasar is observed prior to the re-ionization epoch (e.g.\ Barkana 2001), the fast evolution of the mean absorption in the spectra of quasars at z$>$ 5 suggests that the universe is approaching the reionization epoch at z$\sim$ 6 (Becker \etal\ 2001, Fan \etal\ 2001b, Djorgovski \etal\ 2001).
In a companion paper (Fan et al. 2001c),
we use cosmological simulations to estimate the evolution
of the ionizing background and to constrain the redshift
of reionization (see also Gnedin 2000, Cen \& McDonald 2001,
Lidz et al. 2001).      

We have obtained an optical VLT spectrum of the quasar to 
independently  measure the Gunn-Peterson trough and 
to set tighter limits on the transmitted flux.
We have also obtained a near-IR spectrum,  
to set constraints on the metallicity of the quasar environment and to better estimate the redshift from the CIV emission line, which unlike \lya\ is not affected by intervening absorption.
Finally we obtained a near-IR VLT spectrum of the second most distant quasar SDSS J1306+0356 at z=5.99 (Fan \etal\ 2001b).
We will use the results from this spectrum in the discussion of the metallicity in high redshift quasars.  

The paper is organized as follows: in Section 2 we describe the
observations and present the optical and near-IR spectra of the two
quasars. In Section 3 we discuss the main results obtained from the
optical spectrum: the Gunn-Peterson trough and the limits we can set
from the new data, as well as the region in which
flux is present, and the proximity effect. In Section 4 we derive
the metallicity in the two highest redshift quasars and discuss its
implication for the star formation timescales.  \\ Throughout the
paper we assume a flat, $\Lambda$-dominated universe with $H_{0} = 65$ km
s$^{-1}$ Mpc$^{-1}$, $\Lambda = 0.65$ and $\Omega = 0.35$
\cite{OS95,KT95}.
\section{Observations and results}
\subsection{Optical spectroscopy}

Optical long slit spectroscopic observations were carried out in service mode 
on 28 and 30th of June 2001, 
with the FORS2 spectrograph mounted on VLT/YEPUN on Paranal.
On each night, the quasar SDSS 1030+0524 was observed for a total of 3600 sec
divided into two 30 minute exposures.
The target was placed at different locations along the slit 
for each individual frame, to minimize problems related to flat fielding 
and other possible systematic errors. Weather conditions were clear but not photometric. The seeing was of order  0.8$''$ on both nights.
We used a 1$''$ slit together with the new holographic grism GRISM\_1028z+29
and the order separation filter OG590, which has greatest
sensitivity at $\lambda \approx 8800$\AA, and covers the wavelength
range we are interested in (7900 \AA\ to 9300 \AA)
This grism also offers the highest resolving power, R$\sim$ 2700,
corresponding to a resolution of $\sim 3.1$ \AA, which will allow us to 
cleanly resolve out the night sky, thereby significantly increasing
our signal-to-noise ratio and our ability to subtract the sky.
The dispersion of the spectrum is 0.68 \AA\ per pixel.
Given the rather high airmass of $1.4 < {\rm sec}(z) < 1.9 $ for the observations, the slit was kept at the parallactic angle.

Image reduction (including bias subtraction, flat fielding and 
correction for slit distortion for each individual frame)
was performed with standard IRAF procedures.
The most critical part of the reduction was the sky-background subtraction,
 especially due to the presence of strong sky emission 
lines in the wavelength  region where the flux is very close to zero. 
In particular the  absolute level of flux detection 
changes slightly (by $\sim 1\sigma$) depending 
on the parameters used for sky subtraction.
We therefore assumed that outside the trace of the spectrum 
of the quasar, i.e. 
at other places along the slit, there was no flux.
Therefore after the normal sky subtraction,
we determined the average of the residual sky signal in a 
large region on either side of the  spectrum  and subtracted this single 
value from the entire data frame.

\begin{figure*}
\epsscale{0.9}
\plotone{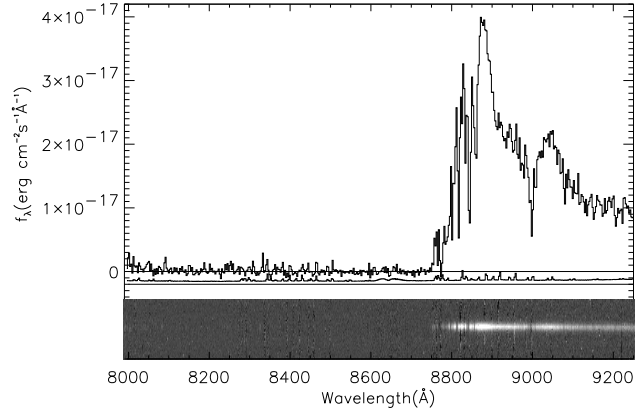}
\caption{Upper panel: the VLT/FORS2 one-dimensional optical spectrum 
of the quasar SDSS 1030+0524  shown in the observed frame and
 smoothed to a dispersion of 3 \AA \ pixel$^{-1}$. 
The bottom line shows the error array derived from the sky spectrum
and offset by $-0.2 \times 10^{-17}$ cgs for clarity. Lower panel: a
grey-scale representation of the sky-subtracted two-dimensional
spectrum plotted on the same wavelength scale. The spatial extent of
the plot is 40 pixels,  corresponding to 20 arcseconds.  
\label{fig1}}
\end{figure*}

Wavelength calibration was done using observations of a Xe-Ar lamp obtained during the morning calibration: the intrinsic accuracy of the wavelength 
fit is better than  0.2 \AA. To this error  we have to add the uncertainty
due to the large range of airmasses of the observations, which we estimate 
from the spectra as less than half a pixel.
The total uncertainty is then 0.4 \AA.
Relative flux calibration was performed using a long slit 
spectrum taken with the same setup of the spectrophotometric 
standard EG274  (Hamuy \etal\ 1992).
The spectra from each night were  calibrated separately and then coadded
using a cosmic-ray rejection algorithm.
Finally the spectrum was extracted: since we could not trace it from
the image itself, due to the absence of intrinsic quasar flux throughout the Gunn-Peterson trough (see below), we used the trace derived from the standard star.
The spectrum was not corrected for telluric absorption.
The two-dimensional spectrum and the extracted one-dimensional spectrum 
(smoothed to 3 \AA per pixel) are presented in Figure \ref{fig1}.
The main characteristics of the Ly$\alpha$ and NV emission lines are given in Table 1.
\subsection{Near-IR Spectroscopy}
Near-IR observations of SDSS 1030+0524 were carried out in service mode 
with ISAAC on VLT-ANTU
on 18 May 2001. Observations were made with the SW Rockwell 1024$\times$1024 pixel Hawaii array which has a scale of 0.148 arcsec/pixel and the lower resolution 
grating, whose resolving power with a 0.6$''$ slit is 860.
The seeing during the night varied between 0.4 $''$ and 0.7$''$.
The total integration time was 5760 sec, split into 2 sets of 16  images. 
We used a detector integration time DIT$=$180 sec over which time a
single integration was done 
(NDIT$=$1). The object was moved between 
exposures by 20$''$ along the slit in a ABBA sequence,
 with an extra jitter of amplitude 5-10$''$ around 
each A and B position.
 The airmass varied between 1.15 and  1.27.

\begin{figure*}
\plotone{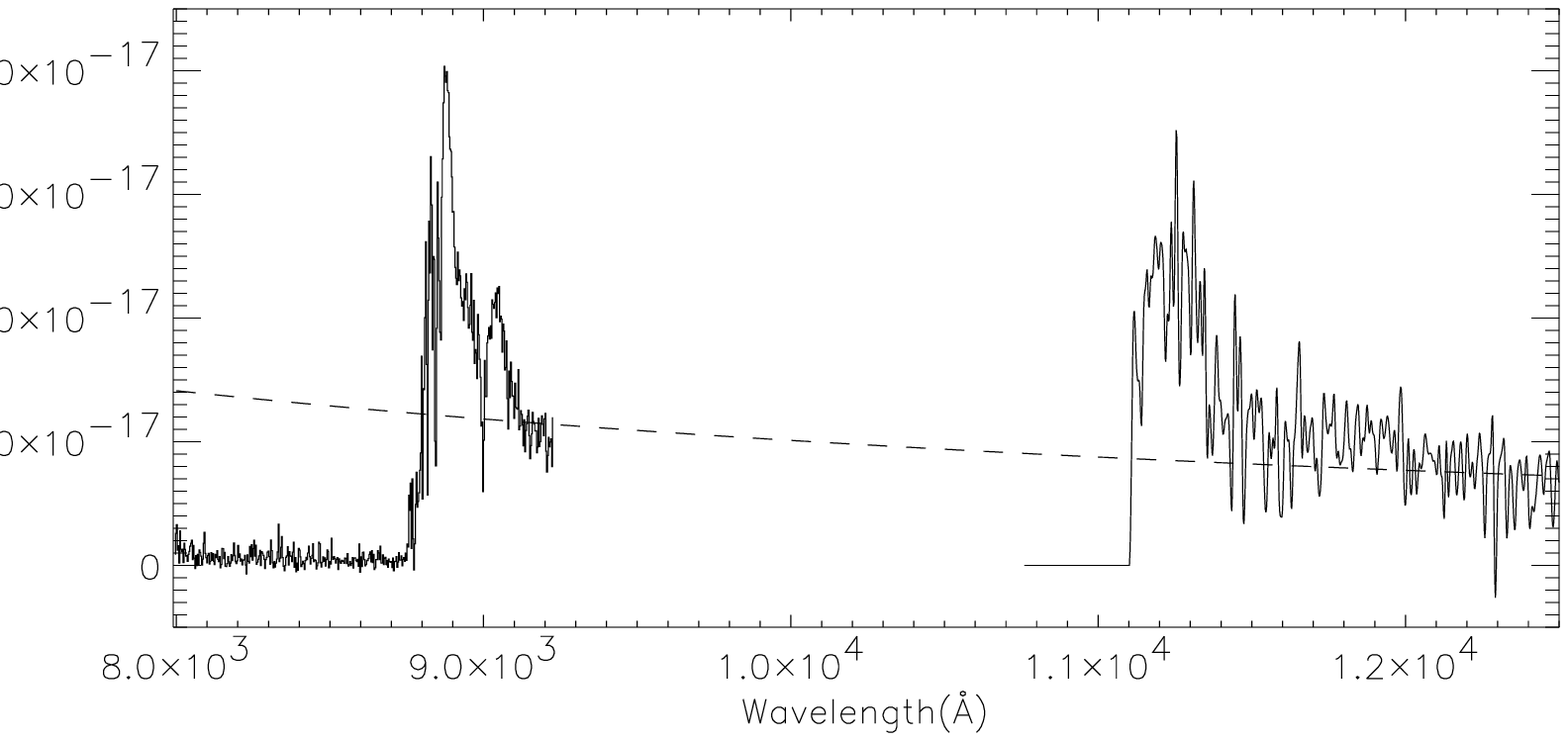}
\caption{A composite of the optical and near IR (J band) spectra of SDSS 1030+0524. The dashed line is the continuum normalized at 12400 \AA.\label{qso1ir} }
\epsscale{0.7}
\plotone{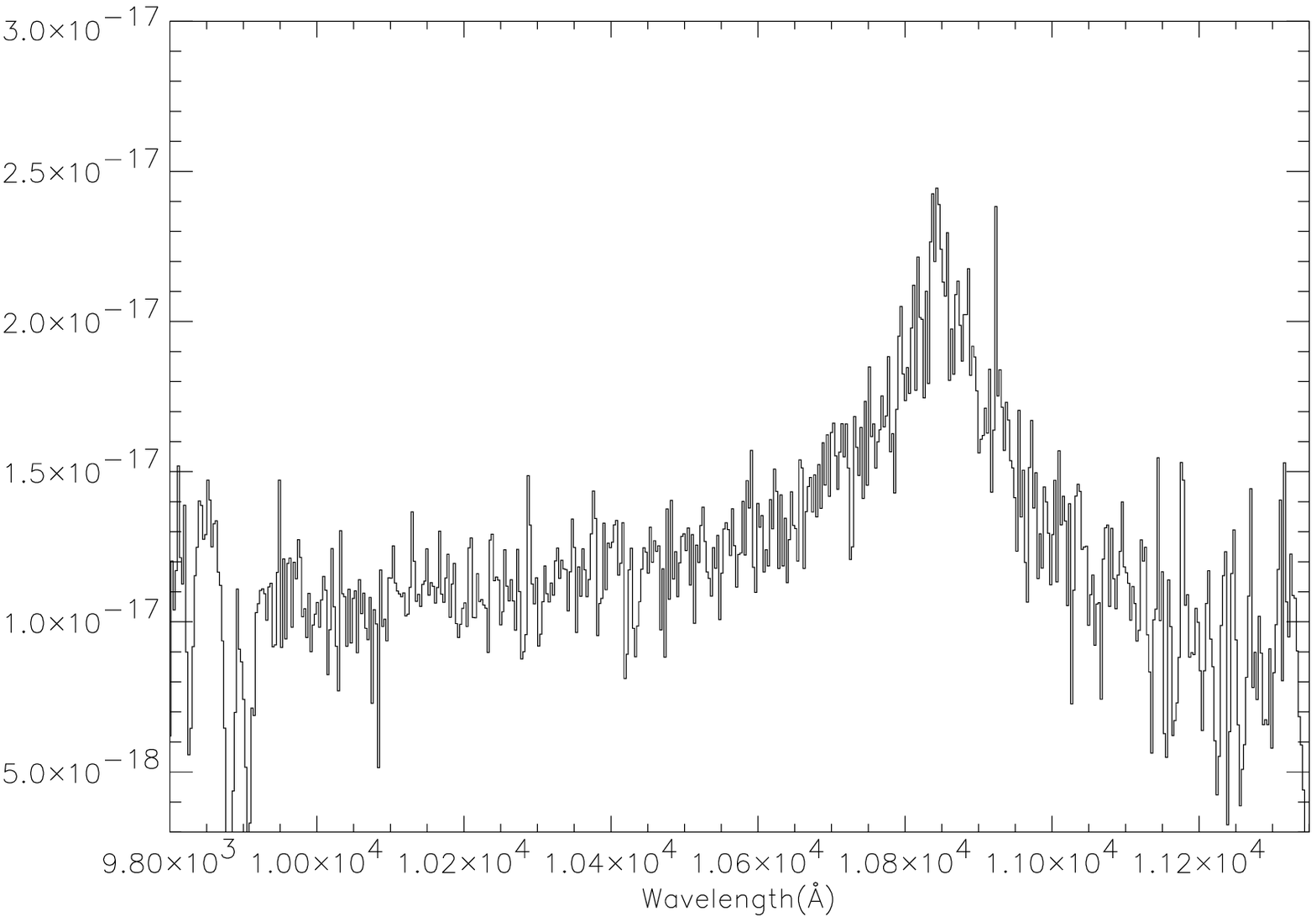}
\caption{ The z-band spectrum of the quasar SDSS 1306+0356, showing the CIV emission line. \label{qso2ir} }
\end{figure*}

The ISAAC data were reduced in the standard way with IRAF, subtracting each image from its adjacent one. After flat fielding, we corrected each 
frame for geometric 
distortions using the sky lines as well as a spectrum of 
an argon lamp, which also led to wavelength calibration.
Finally the frames were aligned using the shifts from  the image 
headers and  combined with the IRAF/avsigclip rejection algorithm.
The spectrum was then extracted using a 3 pixel aperture, 
with a trace derived from the spectrum itself.
Unfortunately the telluric star from the ESO standard calibration plan 
was observed with a different instrument configuration.
Since no telluric spectra were present in the ESO database,
the observations were repeated on the 15th of June with a seeing of around 0.5$''$ and at similar airmass as the science observations.
The standard telluric star  (H 78530) was reduced in a similar way as the spectrum, and in particular the wavelength calibration was done with a lamp exposure taken on the same night. 
However, it is apparent that the removal of telluric features from 
our data is not perfect and the resulting spectrum is 
noisier than we had hoped. 
Finally the spectrum was flux calibrated using the total J band magnitude 
of the object (J$=$18.87$\pm$ 0.10, Fan \etal\ 2001b) and is presented in Figure \ref{qso1ir}. 
The measured characteristics of the CIV line and the 3$\sigma$ limits on the undetected HeII line are reported in Table 1. Contrary to what we hoped, the redshift we derive from the CIV line is actually less accurate than that derived from the lines in the optical spectrum: the reason is the uncertainty in the removal of  some deep telluric features at the same wavelength as the 
CIV emission line. 

 An infrared spectrum of the quasar SDSS J1306+0356 at z$=$5.99 was obtained with ISAAC  with the aim of detecting the CIV line.
The observations were carried out in service mode on 1 July 2001; the total integration time was 5760 sec, which was divided into 2 sets of 16 images, each with 180 sec integration time.
We used the SZ filter, which gives a total wavelength coverage from
0.98 $\mu$m to 1.14$\mu$m. The airmass during the observations varied
from 1.19 to 1.39 and the typical seeing was 0.8$''$.
The observational strategy was the same as for SDSS 1030+0524. Reduction was carried out as described above. The telluric features were eliminated with observation of the telluric standard H74113 , observed with the same set-up of the target during the same night.
The resulting spectrum is shown in Figure \ref{qso2ir}. The two 
absorption features at $\lambda \sim$9900 \AA\  that are observed 
in the spectrum are due to MgII absorption at redshifts 2.20 and 
2.53, respectively.
No obvious intergalactic CIV absorption is seen.

From a fit to the peak of the CIV line we derive a redshift of 6.00$\pm$0.01 
for this quasar, consistent with the z$_{em}$=5.99$\pm$0.02  derived
by Becker \etal\ (2001) from the optical emission lines (NV).
Note that the CIV line is often offset relative to the systemic 
quasar redshift, e.g. for z$\sim 2$ SDSS quasars 
Richards \etal\ (2002) 
find an average shift between the CIV line and the systemic redshift 
on the order of 800 km/s, 
with some shifts as large as 3000 km/s 
(see also Vanden Berk \etal\ 2001, Laor \etal\ 1995).
However in our case the CIV redshift 
agrees within the errors with the redshift derived by NV which is normally very close to the systemic redshift, so we believe there are no such extreme 
offsets; in the error quoted above we have included 
a possible uncertainty of 500 km/s.
The profile of the CIV line actually suggests the presence 
of both broad and narrow 
components; fitting the overall shape with two gaussians gives FWHM
for the two components of 1600$\pm$200 km/s and 4400$\pm$400 km/s
respectively; the broad component is blueshifted relative to the narrow one.
No limits can be derived for the undetected HeII line since it falls at 
the edge of the spectrum where the S/N is very low.

\section{Results}
\subsection{The Gunn-Peterson trough: the limits from the VLT spectrum} 
In the bottom panel of Figure \ref{fig1}, we show the two-dimensional
spectrum of SDSS 1030+0524 in the wavelength region between 7990 \AA \ and 9250 \AA. Clearly there is a sharp drop in flux, blueward of the \lya\ emission, 
and the flux is consistent with zero in the region between 
8450 \AA \ and 8710 \AA.

Following previous papers (e.g. Fan et al. 2001b)
we have computed the transmitted flux ratio over 3 small redshift windows in z$_{abs}$ as $T(z_{abs}) = \langle f_{\nu}^{obs}/f_{\nu}^{con} \rangle$, 
where $f_{\nu}^{obs}$ is the observed flux and  $f_{\nu}^{cont}$
is the intrinsic quasar continuum flux. For the latter quantity we 
have assumed  a power law, with 
$f_{\nu}^{con} \propto f_{\nu}^{-0.5}$ (or 
$f_{\lambda}^{con} \propto f_{\lambda}^{-1.5}$). We normalized this to the 
continuum flux measured at an observed wavelength of 9200 \AA (with the normalization value $f_\lambda=1\times 10^{-17}$ erg s$^{-1}$ cm$^{-2}$ \AA$^{-1}$), since this is the only range of our optical spectrum free from emission and absorption lines. 

The average flux ratio in the three wavelength bins 
is 0.046$\pm$ 0.005, 0.029$\pm$ 0.004 
and 0.001 $\pm$ 0.004  respectively in the intervals 8000-8200 \AA,
8200-8450 \AA \ and 8450-8710 \AA, approximately corresponding to
z$_{abs}$ of 5.65, 5.85 and 6.05. The error bars include only the photon noise, and do not reflect the
systematic error due to the unknown continuum shape (which however should be
smaller, as argued by Fan \etal\ 2001b).  
The results are consistent with those reported by Becker et al. (2001) from the Keck spectroscopic observations. 
Therefore, we confirm the non-detection of flux found in Keck spectrum 
in the same wavelength range. We emphasize that these 
are completely independent measurements, using a different telescope at a
different site with a different instrument, and both spectra show the same 
non-detection of flux.

The implied drop of flux for the wavelength region 8450-8710 \AA\ is a factor 
of 200 at a 1$\sigma$ level, a stronger  
limit than previously reported by Becker et al. (2001), 
who derived a factor of 140, also at 1$\sigma$. 
The corresponding effective Gunn-Peterson optical depth, $\tau_{eff}\equiv -\ln(T)$, is $>$ 5.3. 
Note that  we cannot compute the D$_A$ parameter defined in Oke \& Korycansky (1982) as D$_A \equiv \langle 1 - f_{\nu}^{obs}/f_{\nu}^{con} \rangle$ measured in the region between restframe \lya\ and Ly$\beta$ ($\lambda$=1050 and 1170 \AA) 
and is normally used to describe the decrement of flux below the \lya\ line,  
because the wavelength range covered by our observations is not quite large enough.

The implication for the ionization state of the IGM are discussed in Fan \etal\ (2001c) where it is shown that the observed properties of the IGM at z$\sim$6 are typical of those in the era at the end of the overlap stage of reionization when the individual HII regions merge. Thus this observation suggests that z$\sim$6 marks the end of the reionization epoch.

\subsection{The spectrum below 8450 \AA: IGM absorption between z=5.58 and 5.95}
The extracted spectrum shows some emerging flux below 8450 \AA.
However, from an inspection of the two-dimensional spectrum it appears 
that a large fraction of the flux comes from a few isolated transmission features, while the rest of the spectrum remains consistent with zero flux. 
In other words, a simple description of the transmitted flux  in this  region 
as arising from a uniform IGM is not sufficient.
The presence of these gaps of transmitted flux can be explained, 
in the context of reionization, as due to cosmological HII regions around the sources that reionize the IGM before the epoch of overlap. 
The characteristics of these gaps, 
such as their sizes and probability  distribution  
are important since they can put constraints 
e.g. on the nature of the  ionizing sources (quasars and/or galaxies) on 
their lifetimes etc.
In particular the reader is referred to Miralda-Escud\'e \etal\ (2000) 
for a detailed discussion of gaps in the 
Gunn-Peterson trough due to individual HII regions 
(see also Loeb \& Barkana 2001,
 Djorgovski \etal\ 2001, Haiman \& Loeb 1999).

\begin{figure*}
\epsscale{1.}
\plottwo{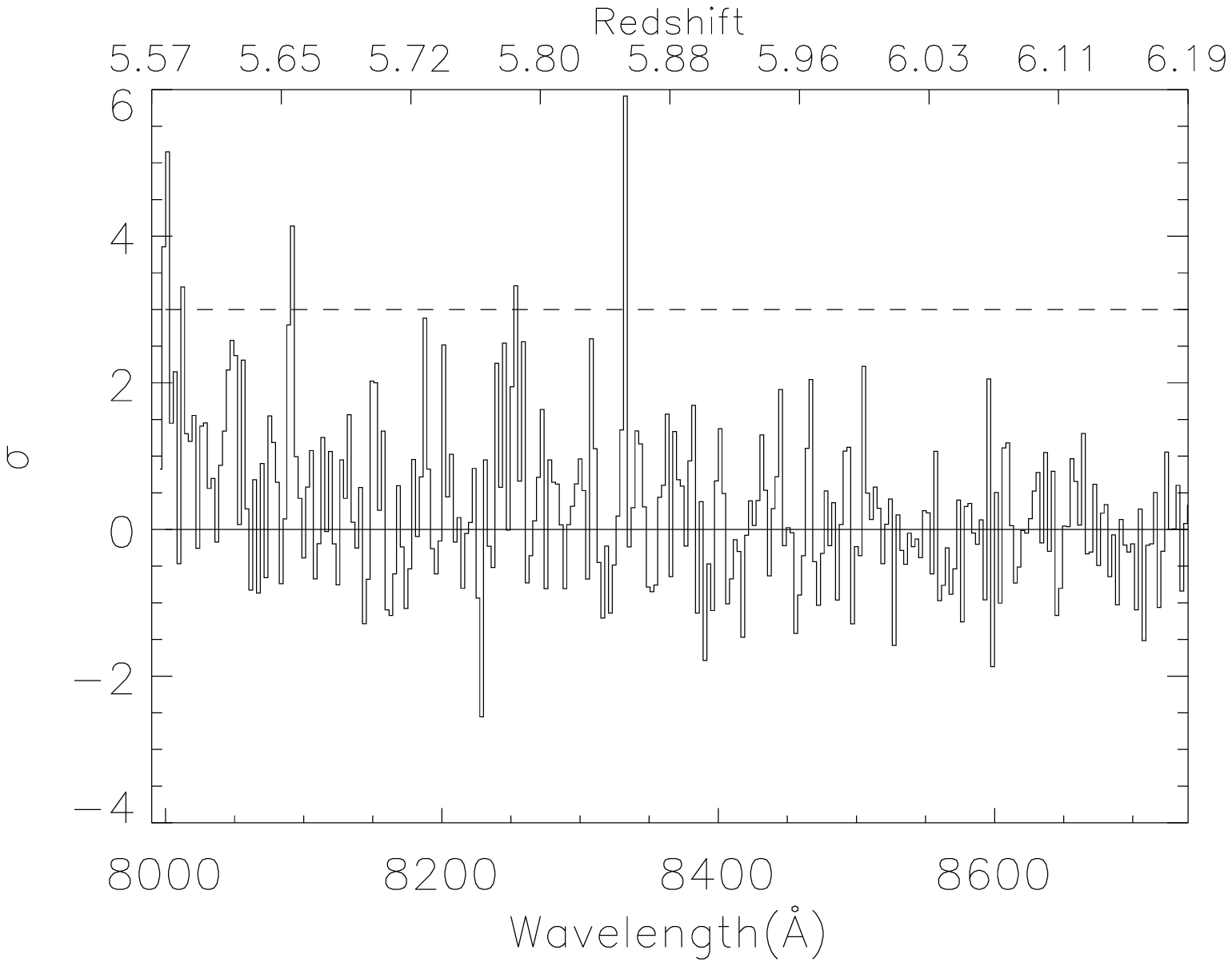}{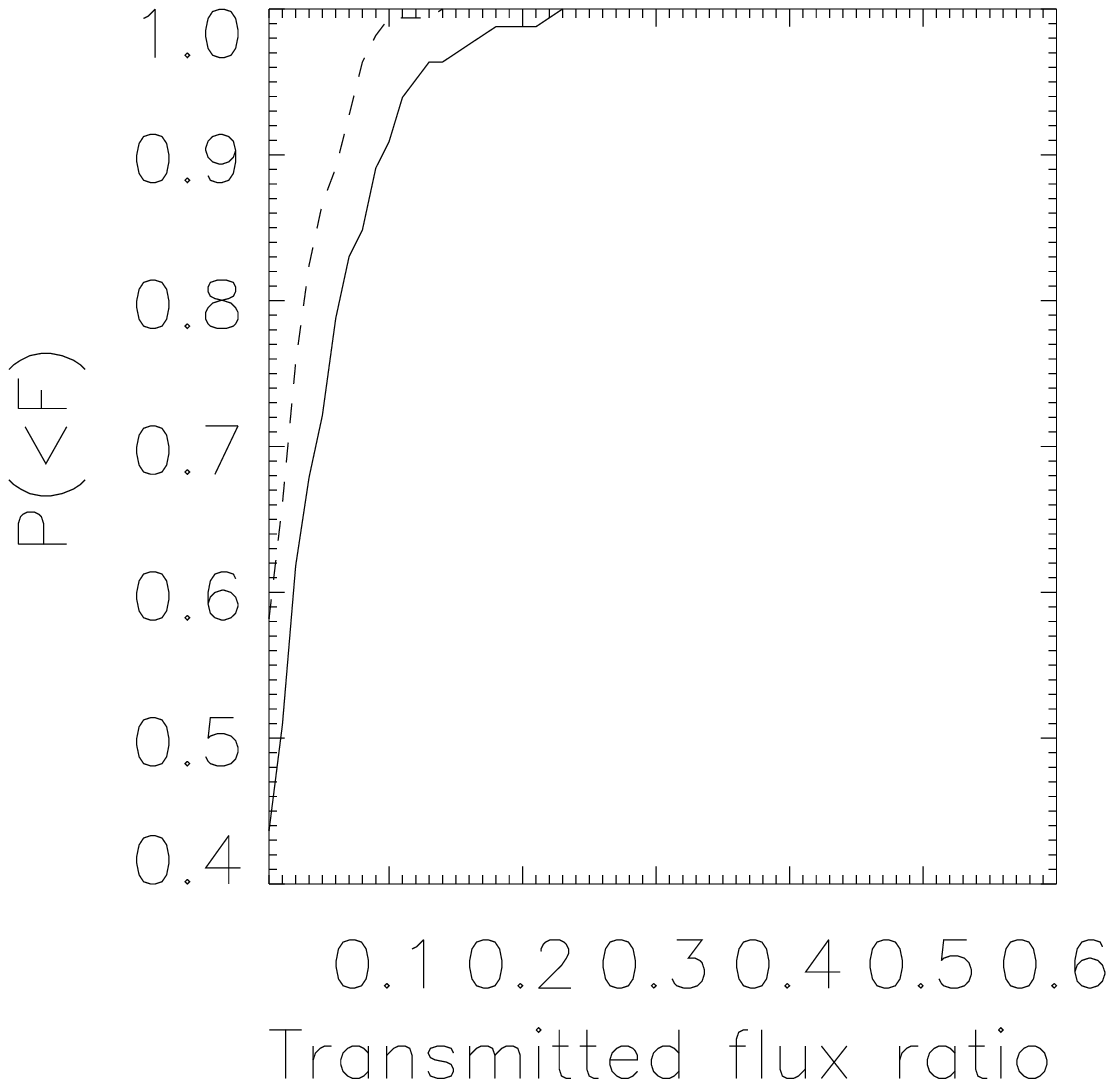}
\caption{Left: an enlargement of the region between 8000 \AA \ and 8750 \AA, plotted in terms of significance, i.e. at each pixel the observed spectrum has been divided by the error array. The dashed line represent the 3 $\sigma$ level: clearly there are no peaks beyond 8400 \AA. 
The spectrum was binned to the resolution ($\sim$ 3 \AA).  
Right: the cumulative distribution of the transmitted flux ratio for the spectrum in the 8000-8400 \AA \ region. The spectrum has been rebinned to the resolution $\sim$ 3\AA/pixel. The dashed line indicates the same distribution for a random realization of a spectrum with zero average flux and the same average noise of the real data. \label{fig2}}
\end{figure*}

In Figure \ref{fig2} (left panel) we show the significance of any 
possible flux detection in the region between 8000 \AA\ and 8750 \AA: for each resolution element (the spectrum was binned to a resolution of 3 \AA \ per pixel) the observed flux has been divided by its corresponding error array.
There are no significant  features ($>$ 3$\sigma$) in the range between 8400-8750 \AA, while there are several such features present in the shorter wavelength part of the spectrum.
All the features above 3$\sigma$ are also clearly recognized 
in the two-dimensional spectrum, and are visible on both spectra from the 
two different nights.
We have tried to quantify what fraction of the total flux below 8400 \AA\ is 
accounted for by these isolated features. 
To detect features in a systematic way, we used SEXtractor (Bertin \& Arnouts 1996) on the two-dimensional sky-subtracted spectrum. 
We considered as an emission feature anything 
 detected in more than 3 adjacent pixels 
with a flux larger than 1$\sigma$ 
plus the local background as determined in an annulus around the feature.
We considered only those features that have a peak of emission along 
the trace of the spectrum ($\pm$ 1 pixel); furthermore we 
eliminated all those features that were coincident with sky lines 
(by using the 
error array of the extracted spectrum), since these could be due to residuals of sky subtraction.
This is a conservative approach since we might  eliminate true features.  
Eight  such features are detected from SEXtractor.
We then measured the flux inside the features, assuming the 
width as given by SEXtractor in the dispersion direction, and 
3 pixels in the spatial direction. The values measured 
are reported in Table 2, together with the central wavelength 
and the width. The flux from these features accounts
for  60\% of the total flux in the 8000-8400 \AA\ range.

To allow comparison with simulations, we also present the
cumulative distribution of the transmitted flux ratio of the region between 
8000 and 8450 \AA\ (corresponding to a mean redshift of absorption  
z$_{abs}=$5.74) in Figure \ref{fig2}. For this plot the spectrum was rebinned to the resolution of 3 \AA/pixel, and then divided by the assumed unabsorbed 
continuum flux estimated in the previous section.
Ninety-nine per cent of the spectrum 
has a transmitted flux  ratio less than  20\% of the continuum, 
and the distribution goes to 100\% very rapidly (note that all 
results depend on the resolution of our spectrum). 
For comparison, we also draw the same distribution for a random
realization of a spectrum having the same resolution and the same average
noise as the real data, but with zero average flux. 
One should actually do this pixel-to pixel using the individual 
pixel errors (see Fan et al.\ 2001c) but such a detailed analysis is 
beyond the scope of this paper.  
In any case the ``random" line is clearly above the data, implying that the observed spectrum has finite non-zero average flux in this wavelength region
and cannot be solely due to the noise alone.
 McDonald \& Miralda-Escud\'e (2001) present the detailed flux distribution from simulated spectra  at slightly lower absorption redshift. 
Their prediction could be easily compared with these data. For example at 
z$_{abs}=$5.2 they predict that only 3\% of the spectrum should have a flux higher than 50\% of the continuum, and less than 0.5\% should have a flux greater than 75\% of the continuum.
Simulations show that because the IGM is highly clumpy,
most of the flux that we observe is transmitted from a few of the most
underdense voids, consistent with what is seen here (see the detailed
discussion in Fan et al. 2001c).                  

\subsection{The proximity effect}

The spectrum shows excess transmitted flux
 immediately blueward of the Ly$\alpha$ line,
as well as several absorption lines.
This is due to the fact that  luminous quasars
such as SDSS 1030+0524  ionize the surrounding regions
and create HII regions of radius several Mpc, the so-called
``proximity'' effect (e.g. Carswell \etal\ 1982).
The transmitted flux extends over about 100 \AA\ blueward (i.e., to
8750 \AA) of the \lya\
emission peak at 8850 \AA.
This corresponds to a distance of about 23h$^{-1}$ Mpc comoving,
or a proper size of D$_{obs}=$3.1h$^{-1}$ Mpc at the epoch z$=$6.28.
 
The size of the HII region associated with the quasar depends on the
luminosity and lifetime of the quasar and the clumpiness of the gas
near the quasar.
As pointed out by Madau \& Rees (2000), if the lifetime of the quasar
is shorter
than the recombination time 
recombination can be neglected, and the evolution of the HII
region can be decoupled from the expansion of the universe.
This holds at z$=$6.28 on Mpc scales
if the clumpiness factor $C$ is less than $\sim$10.
The photons will travel freely in the highly ionized bubble and
will be absorbed in a transition layer (I-front).
This I-front will initially expand at a velocity very close to the speed
 of light, but the Ly$\alpha$ and continuum photons  from the quasar  cannot
``catch up'' with the boundary of the HII region until its expansion
speed slows down (Cen \& Haiman 2000), when it first begins to be
limited by the number of ionizing photons. 

In practise, the volume of ionized region that we observe will be 
proportional to the number of Lyman-continuum photons that are 
emitted over the source lifetime.
In a  homogeneous medium we can write: 
$$V_I= {4\over 3} \pi D_I^3 = {t_Q \dot{N}_{ph}\over n_H f}$$
where $\dot{N}_{ph}$
is the photo-ionization rate, and $n_H$ is the mean hydrogen
density within $D_I$.
Following the papers cited, and in particular Cen \& Haiman 2000, we use the mean hydrogen density  
given by $n_H = 1.6 \times 10^{-7} (1+z)^3 (\Omega_b h^2/0.02) $
cm$^{-3}$. 
However as we remarked earlier, luminous quasars probably reside
in significantly overdense regions: the unknown bias factor will contribute to the uncertainty of the proximity effect region.
In the equation above, $f$ is the fraction of neutral hydrogen present 
when the quasar switches on.
From the observations the only constraint  we  have is 
that  $f \ge 0.01$ at $z\sim6$ (Fan \etal\ 2001c).
Simulations (e.g.\ Gnedin 2000)
indicate that the ionizing fraction could be $f=0.1$ or more at $z=6.28$. 
Taking the quasar luminosity as $L_\nu \propto \nu^{-\alpha}$, we 
derive a total photo-ionization rate for the quasar of 
$\dot{N}_{ph}=2/\alpha \times 10^{57}$ s$^{-1}$.
Therefore the quasar must have been on for 
$t_Q \sim 1.3 f \alpha \times 10^7$ yrs. This 
is the total time during which the quasar has been emitting ionizing photons,
and not necessarily the time elapsed since it first turned on.
The derived time is of the same order as the
e-folding timescale for a black hole accreting with 0.1 efficiency of the
 Eddington luminosity of  4$\times$ 10$^7$ yr (Salpeter 1964).

Note that $t_Q$ is shorter than the 
light-crossing time of the HII region ($\sim$1.5 $\times 10^{7}$ yrs). 
The reason is that photons along the line of sight have been emitted 
at different redshift and therefore at different lookback times. 
Therefore it is not correct to set a minimum 
lifetime of the quasar by saying that those photons that are 
ionizing  the edges of the HII region 
needed a certain time to propagate there, travelling at the speed c, 
since they could have been emitted at a later time.
The light-travel  argument would only be right if we could observe 
the HII region seen around the quasar in the plane of the sky.

The above calculation does not take into account other effects
such as the presence of Lyman limit systems near the quasar,
which would completely absorb the flux and produce a cutoff in
the spectrum. In fact such  a single overdense (and therefore neutral)
cloud along the line of
sight between the quasar and us is quite likely to exist close
enough to the quasar to give a biased
answer for the size of the ionized region (see Fan \etal\ 2001c).
The presence of a Lyman limit system close to the quasar would lead
to an underestimate of the quasar lifetime.

The transmitted flux profile blueward of the \lya\ line also
shows numerous absorption features which are produced by the density fluctuations of the medium around the quasar but are not resolved in the present 
spectrum. Cen \& Haiman (2000)
show that much higher spectral resolution measurements  (R$\sim$10$^4$)
of the blue side of \lya\ in bright quasars
would provide direct estimates of the density
fluctuations of gas on small scales.

The proximity effect is one of the primary techniques used to estimate the
photo-ionization rate of the IGM  (e.g. Bajtlik, Duncan, \& Ostriker 1988, Bechtold 1994), by
determining the distance from the quasar at which the optical depth of
the Ly$\alpha$ forest is half that in the Gunn-Peterson trough.
However, in the case of our quasar, the ionizing background is
dominated by the flux of the quasar itself, so the profile
of the \lya\ line is insensitive to the ionization state of the IGM.  

\begin{figure*}
\epsscale{0.5}
\plotone{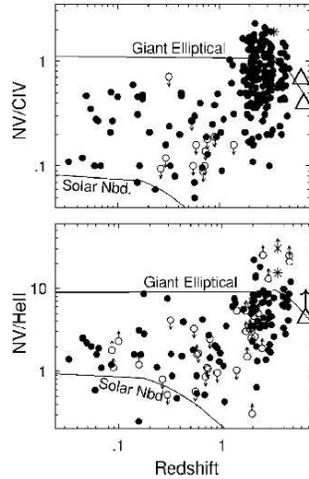}
\caption{Measured NV/CIV and NV/HeII line ratios versus redshift 
for quasars up to z$\sim$ 4 reproduced from Hamann and Ferland (1999) and 
with the addition of the new values measured for the 2 highest redshift 
quasars (represented as triangles).
 \label{metal}}
\end{figure*}

\section{Metallicity in the two highest redshift quasars}
The metallicity ($Z$) of the gas around high-redshift quasars contains 
important
information on the enrichment history of the gas in the quasar
environment, and probes the star formation that came before the epoch of the 
observations,
possibly the first stars that formed in massive collapsed structures
(e.g.\ Hamann \& Ferland 1999). 
In particular, line ratios involving nitrogen are valuable 
tracers of chemical enrichment, since this element is selectively  
enhanced by secondary processing in stellar populations 
and its abundance increases roughly as Z$^2$.

There is increasing evidence from measurements of various line ratios that 
the metallicity of the broad line regions around quasars have solar or 
supersolar values even at redshifts higher than 4, implying that there are 
stars at that redshift with ages exceeding 1 Gyr (e.g.\ Elston \etal\ 1994, 
Thompson \etal\ 1999). 
Clearly, dating the first star formation becomes increasingly 
interesting at redshifts of 5 or beyond, as the age of the universe 
approaches the enrichment timescale itself.
At redshifts 5.99 and 6.28 the universe is  only 910 Myr and 860 Myr old 
respectively, in our adopted cosmology.

From our spectrum of the z$=$6.28 quasar,
 we are able to measure the characteristics of the Ly$\alpha$, NV and CIV emission lines,  which are reported in Table 1. We do not detect HeII, but we can set an upper limit based on the  S/N of the continuum. For the  redshift 5.99 quasar, we combine our measurement of the CIV line with  the Ly$\alpha$ and NV values derived by Becker \etal\ (2001).
The derived line ratios for the redshift 6.28 quasar are  NV/CIV$=$ 0.35, slightly lower but consistent with that previously reported by Becker \etal\ (2001), and NV/HeII $>$ 4.3.
For the redshift 5.99 quasar, the restframe EW of the CIV line is
37$\pm$ 6 \AA; using the reported restframe EW of NV from Becker \etal\ (EW$=$17.9 \AA), we derive for this object a  NV/CIV ratio of 0.67.
Note that in all cases we have derived the line ratios using the EW of each line and assuming that the continuum has the form of a power law 
with  $f_{\nu} \propto \nu^{-\alpha}$ and $\alpha=0.5$ as previously used.

In Figure \ref{metal} we show the measured values for 
NV/HeII and NV/CIV, in a plot of line ratios versus redshift 
reproduced from  Hamann \& Ferland (1999), which includes measured 
values for quasars at different redshifts.
The line ratios of z$\sim$6 quasars appear fully consistent with those at 
redshift between 2 and 4, indicating little evolution in the abundances, and hence metallicities, of quasar environments.
However metallicity might be actually more a function of local density 
than of redshift (e.g. Cen \& Ostriker 1999); we also know  
that metallicity scales with quasar luminosity, 
and since the two quasars are the brightest objects at that redshift, 
we might argue that 
the most luminous objects at z$\sim 6$ have lower
 metallicity than their counterparts at z$\sim$ 4. 
Clearly more statistics are needed to determine whether we are starting to 
see a decline in metallicity as compared to z$\sim$4. 

Figure \ref{metal} also shows the 
predictions based on the Hamann \& Ferland  (1993,1999) chemical evolution models.
The models that best reproduce the observed ratios for the two z$\geq$ 6 
quasars are  those of ``Giant Elliptical" nitrogen enrichment; these models  
involve  short star formation timescales (0.5 Gyr) and IMF favoring high mass stars.
The solar neighborhood models severely under-predict the two line ratios (NV/HeII and NV/CIV). 
If we plot the observed line ratios into Figure 6 of 
Hamann \& Ferland (1999), we derive a metallicity 
of roughly 7-8 $Z_{\odot}$ for our two quasars. 
We emphasize that the actual value of the metallicity is somewhat uncertain, 
due to the uncertainty in deriving reliable abundances from the 
emission line ratios: e.g. varying the shape of the EUV
continuum, varying the mix of ionization parameter and column density in
the radiating regions, and allowing for scattering contributions to NV 1240
(e.g. Krolik \& Voit, 1998) 
can all perturb the relative line strengths without indicating
anything about the abundances.
However the main result is that
the observed ratios definitely require $Z \geq Z_{\odot}$.
Note that metallicity can be derived 
self-consistently from the two separate line ratios only if N$\propto Z^2$. 

From the  models of Hamann \& Ferland (1993) we see that the
  minimum time needed for a stellar population to achieve 
the observed metallicity in any of their models is about 300 Myr.
In the assumed cosmology the implied redshift of star formation is z$\sim$8.7, 
an epoch when the universe was about $\sim$560 Myr old. 
Even if the quasar environments had only solar metallicity, the time required 
for any stellar population to reach such a metallicity would be on the 
order of 100 Myrs or more, implying that the star
    formation must have begun well before the observed quasar activity. 
Again these results are effected by the  uncertainties of the models 
outlined above: other elements and in particular Fe, 
might provide better ``clocks" 
for constraining the age of quasars 
and the epoch of first star formation. Fe has a large delayed 
contribution from type Ia supernovae, with a predicted time delay of
 about 1 Gyr after the initial starburst 
(Greggio \& Renzini 1983, Matteucci \& Greggio 1986) which does not depend
 on other parameters such as star formation rate etc. Therefore the ratio of 
Fe to $\alpha$ elements such as Mg  
is an absolute clock for constraining the ages of star forming regions. 
This will be pursued with follow-up observations in K-band where 
there are interesting features such as the broad FeII$\lambda$2960 and the 
MgII$\lambda$2798 lines.
\section{Conclusion}
We have presented new optical and near-IR spectroscopic observations of the 
two most distant quasars known, at redshifts 6.28 and 6.0. 
We have confirmed the presence of a complete GP trough in the highest 
redshift object, derived new limits on the transmitted flux
and optical depth in this region, and analyzed the characteristics of 
the flux emission in the region.
In Fan et al. (2001c) we 
use semi-analytic models to study the evolution of the ionizing background 
and the epoch of reionization from the z$\sim$ 6 quasars.

From the optical and near-IR spectra, we have estimated the metallicities 
of these early quasar environments, finding supersolar values for both objects.
These high metallicities 
imply that the first stars around the quasars must have formed at least a few hundreds of Mpc prior to the observation, i.e. at redshifts higher than 8.  
Follow-up observations of other metal emission lines will help us to 
better constrain the ages of the star forming regions.



\acknowledgements

The Sloan Digital Sky Survey (SDSS) is a joint project
of The University of Chicago, Fermilab, the Institute for
Advanced Study, the Japan Participation Group,
The Johns Hopkins University, the Max-Planck-Institute for
Astronomy (MPIA), the Max-Planck-Institute for Astrophysics (MPA),
New Mexico State University,
Princeton University, the United States Naval Observatory,
and the University of Washington. Apache Point
Observatory, site of the SDSS telescopes,
is operated by the Astrophysical Research Consortium (ARC).
Funding for the project has been provided by the
Alfred P. Sloan Foundation, the SDSS member institutions,
the National Aeronautics and Space Administration,
the National Science Foundation, the U.S. Department of
Energy, the Japanese Monbukagakusho,
and the Max Planck Society. The SDSS Web site is
http://www.sdss.org.
Michael Strauss acknowledges the support of NSF grant AST-0071091.




\clearpage

\begin{table}
\begin{center}
\caption{Emission line properties.\label{tbl-1}}
\begin{tabular}{cllll}
\tableline\tableline 
Line       &  redshift     & flux          & FWHM        & EW\tablenotemark{a}  \\
           &               &  10$^{-15}$erg s$^{-1}$ cm$^{-2}$ & km s$^{-1}$ & \AA \\
\tableline  
Ly$\alpha$ & --            & 2.70          & 4000$\pm$400&41.9$\pm$4     \\
NV         & 6.29 $\pm$0.01& 0.86          & 2400$\pm$260   & 14.3$\pm$4    \\
CIV        & 6.27 $\pm$0.02&  3.2          & 5600$\pm$400   & 42.5$\pm 5$  \\
HeII       & --            &  $<$0.20      & ---         & $<$ 3.3     \\
\tableline
\end{tabular}
\tablenotetext{a}{Restframe EW}
 
\end{center}
\end{table}

\begin{table}
\begin{center}
\caption{Characteristics of emission features in the 8000-8400 \AA\  range of the spectrum.\label{tbl-2}}
\begin{tabular}{cccc}
\tableline\tableline                   
num & $\lambda_0$  & Flux    &  Width(${1\sigma}$)  \\ 
    & \AA          & 10$^{-17}$erg s$^{-1}$ cm$^{-2}$ & \AA \\
\tableline    
1 &  8000.4                 & 1.59         & 8.2\\ 
2 &  8006.3                 & 0.43          & 2. \\ 
3 &  8012.7                 & 0.92          & 4.8\\
4 &  8047.1                 & 1.02          & 7.5\\
5 &  8091.4                 & 1.12          & 5.4\\
6 &  8150.6                 & 0.49          & 4.1\\
7 &  8253.1                 & 0.67          & 4.8\\
8 &  8332.9                 & 1.02          & 4.8\\
\tableline
\end{tabular}

\end{center}
\end{table}

\end{document}